\documentclass[11pt,a4paper,notitlepage]{article}

\usepackage[margin=1in]{geometry}
\usepackage[utf8]{inputenc}
\usepackage[T1]{fontenc}
\usepackage{times}
\usepackage{hyperref}
\usepackage{url}
\usepackage{booktabs}
\usepackage{amsfonts}
\usepackage{amsmath}
\usepackage{microtype}
\usepackage{graphicx}
\usepackage{multirow}
\usepackage{xcolor}
\usepackage{natbib}
\usepackage{titlesec}
\usepackage{abstract}

\titlespacing*{\section}{0pt}{10pt}{4pt}
\titlespacing*{\subsection}{0pt}{8pt}{3pt}
\titlespacing*{\paragraph}{0pt}{6pt}{4pt}

\hypersetup{
    colorlinks=true,
    linkcolor=blue,
    citecolor=blue,
    urlcolor=blue
}

\begin{document}

\noindent\rule{\textwidth}{1.5pt}
\vspace{2pt}
\begin{center}
  {\LARGE \textsc{StratRAG: A Multi-Hop Retrieval Evaluation Dataset\\[4pt]
  for Retrieval-Augmented Generation Systems}}
\end{center}
\vspace{2pt}
\noindent\rule{\textwidth}{0.8pt}

\vspace{14pt}

\begin{center}
  {\large \textbf{Aryan Patodiya}}\\[5pt]
  Department of Computer Science\\
  California State University, Fresno\\
  Fresno, CA, USA\\[3pt]
  \texttt{aryanpatodiya018@gmail.com} \quad
  \href{https://huggingface.co/datasets/Aryanp088/StratRAG}
       {\texttt{hf.co/datasets/Aryanp088/StratRAG}}
\end{center}

\vspace{14pt}

% ── Abstract ──────────────────────────────────────────────────
\renewcommand{\abstractname}{\normalfont\large\bfseries\scshape Abstract}
\begin{abstract}
I introduce \textbf{StratRAG}, an open-source retrieval evaluation
dataset designed to benchmark Retrieval-Augmented Generation (RAG)
systems on multi-hop reasoning tasks under realistic, noisy document-pool conditions.
Derived from HotpotQA \citep{yang2018hotpotqa} (distractor setting),
StratRAG comprises 2{,}200 examples across three question types ---
\textit{bridge}, \textit{comparison}, and \textit{yes-no} --- each
paired with a pool of 15 candidate documents containing exactly 2
gold documents and 13 topically related distractors.
I benchmark three retrieval strategies --- BM25, dense retrieval
(MiniLM-L6-v2), and hybrid fusion --- reporting Recall@$k$, MRR,
and NDCG@5.
Hybrid retrieval achieves the best overall performance
(Recall@2\,=\,0.70, MRR\,=\,0.93), yet bridge questions remain
substantially harder (Recall@2\,=\,0.67), motivating future work on
reinforcement-learning--based retrieval policies.
StratRAG is publicly available at
\url{https://huggingface.co/datasets/Aryanp088/StratRAG}.
\end{abstract}


% ── 1. Introduction ───────────────────────────────────────────
\section{Introduction}

Retrieval-Augmented Generation (RAG) has emerged as a dominant
paradigm for grounding large language model (LLM) outputs in
verifiable external knowledge \citep{lewis2020rag}.
A RAG pipeline consists of two stages: a \textit{retriever} that
selects relevant documents from a corpus, and a \textit{generator}
that produces an answer conditioned on those documents.
While significant effort has been devoted to evaluating generation
quality, retrieval quality under realistic, noisy conditions remains
comparatively underexplored.

Existing multi-hop QA benchmarks such as HotpotQA
\citep{yang2018hotpotqa}, MuSiQue \citep{trivedi2022musique}, and
2WikiMultiHopQA \citep{ho2020constructing} provide rich question
sets but are not structured for direct retrieval evaluation: they lack fixed-size document pools, do not guarantee gold-document positions, and do not provide controlled distractor sets for benchmarking.

I address this gap with \textbf{StratRAG}, a dataset that
restructures HotpotQA into a retrieval-first evaluation format.
Each example in StratRAG presents a retriever with a pool of 15
documents --- 2 gold and 13 distractors --- and asks it to identify
the gold documents given only the question.
This design enables direct computation of retrieval metrics
(Recall@$k$, MRR, NDCG@$k$) independently of any downstream
generator, making it suitable for ablation studies, retriever
development, and RAG system evaluation.

My contributions are as follows:
\begin{itemize}
  \item I release StratRAG, a clean, structured retrieval benchmark
        derived from HotpotQA with verified gold-document indices across all 2{,}200 examples.
  \item I provide baseline results for three retrieval strategies
        (BM25, dense, hybrid) across five standard metrics.
  \item I identify bridge questions as the primary open challenge,
        motivating research into RL-based retrieval policies for multi-hop reasoning.
\end{itemize}

% ── 2. Related Work ───────────────────────────────────────────
\section{Related Work}

\paragraph{Multi-hop QA datasets.}
HotpotQA \citep{yang2018hotpotqa} introduced the distractor setting
in which 8 distractor paragraphs accompany 2 gold paragraphs,
requiring models to reason across documents.
MuSiQue \citep{trivedi2022musique} introduced more compositional
multi-hop chains, and 2WikiMultiHopQA \citep{ho2020constructing}
leveraged structured Wikipedia relationships.
StratRAG differs from these in its explicit focus on retrieval
evaluation: gold documents are always at known positions, pool size
is fixed, and all empty-text paragraphs are filtered out.

\paragraph{RAG evaluation.}
RAGAS \citep{es2023ragas} proposes end-to-end RAG evaluation metrics
including faithfulness and answer relevance, but does not isolate the
retrieval step.
BEIR \citep{thakur2021beir} is a widely used retrieval benchmark
covering diverse domains, but does not target multi-hop reasoning
specifically.
StratRAG complements these by providing a focused, multi-hop
retrieval benchmark with controlled document pools.

\paragraph{Retrieval methods.}
BM25 \citep{robertson2009bm25} remains a strong sparse retrieval
baseline.
Dense retrieval using bi-encoders \citep{karpukhin2020dpr} has
shown strong performance on single-hop QA.
Hybrid methods combining sparse and dense scores
\citep{ma2022hybrid} consistently outperform either alone on
heterogeneous benchmarks.
I evaluate all three approaches on StratRAG.

% ── 3. Dataset Construction ───────────────────────────────────
\section{Dataset Construction}

\subsection{Source Data}
StratRAG is derived from HotpotQA \citep{yang2018hotpotqa} in its
distractor configuration, which provides 10 paragraphs per question
(2 gold, 8 distractors) from Wikipedia.
I use the official \texttt{train} and \texttt{validation} splits.

\subsection{Processing Pipeline}
For each HotpotQA example, I apply the following steps:

\begin{enumerate}
  \item \textbf{Gold document identification.}
        Supporting fact titles from the \texttt{supporting\_facts}
        field are matched against context paragraph titles (after
        Unicode NFC normalization) to identify gold paragraphs.

  \item \textbf{Empty paragraph filtering.}
        Distractor paragraphs with no sentence content are excluded.
        HotpotQA contains a non-trivial number of such paragraphs;
        retaining them would introduce spurious empty-text documents
        into the pool.

  \item \textbf{Document pool construction.}
        Gold documents are placed at indices 0 and 1 of \texttt{doc\_pool}.
        Remaining slots (up to a fixed pool size of 15) are filled with randomly shuffled valid distractor paragraphs.

  \item \textbf{Question type labelling.}
        Each question is assigned one of three types:
        \textit{yes-no} (answer is ``yes'' or ``no''),
        \textit{comparison} (question contains comparative keywords),
        or \textit{bridge} (all others).
\end{enumerate}

\subsection{Dataset Statistics}

Table~\ref{tab:stats} summarizes StratRAG's key statistics.

\begin{table}[h]
\centering
\caption{StratRAG dataset statistics.}
\label{tab:stats}
\begin{tabular}{lrr}
\toprule
\textbf{Statistic} & \textbf{Train} & \textbf{Validation} \\
\midrule
Total examples        & 2{,}000 & 200  \\
Bridge questions      & 1{,}775 & 171  \\
Comparison questions  &     112 &  17  \\
Yes-no questions      &     113 &  12  \\
Docs per pool         &      15 &  15  \\
Gold docs per example &       2 &   2  \\
Distractor docs       &      13 &  13  \\
Empty gold indices    &       0 &   0  \\
\bottomrule
\end{tabular}
\end{table}

\subsection{Data Format}
Each example is a JSON object with the following fields:
\texttt{id}, \texttt{query}, \texttt{reference\_answer},
\texttt{doc\_pool} (list of 15 dicts with \texttt{doc\_id},
\texttt{text}, \texttt{source}), \texttt{gold\_doc\_indices},
\texttt{metadata} (\texttt{split}, \texttt{question\_type}),
\texttt{created\_at}, and \texttt{provenance}.

% ── 4. Benchmark Experiments ──────────────────────────────────
\section{Benchmark Experiments}

\subsection{Retrieval Methods}

I evaluate four retrieval conditions on the StratRAG validation set
($n = 200$):

\paragraph{Random.}
Documents are shuffled uniformly at random.
This provides a theoretical lower bound; expected Recall@2
$= 2/15 \approx 0.133$.

\paragraph{BM25.}
I use Okapi BM25 \citep{robertson2009bm25} implemented via
\texttt{rank\_bm25}.
Both query and document texts are lowercased and whitespace-tokenized.

\paragraph{Dense.}
I encode queries and documents using
\texttt{all-MiniLM-L6-v2} \citep{wang2020minilm} from the
Sentence Transformers library \citep{reimers2019sentencebert}.
Retrieval scores are cosine similarities between query and document
embeddings.

\paragraph{Hybrid.}
BM25 and dense scores are each min-max normalized to $[0, 1]$ and
combined with equal weight ($\alpha = 0.5$):
\[
  s_{\text{hybrid}}(d, q) =
    \alpha \cdot s_{\text{BM25}}(d, q) +
    (1 - \alpha) \cdot s_{\text{dense}}(d, q)
\]

\subsection{Metrics}
I report Recall@$k$ ($k \in \{1, 2, 5\}$), Mean Reciprocal Rank
(MRR), and NDCG@5.
Recall@$k$ measures the fraction of gold documents found in the top
$k$ retrieved documents.
MRR measures the reciprocal rank of the first gold document.
NDCG@5 is the normalized discounted cumulative gain at rank 5.

\subsection{Results}

Table~\ref{tab:results} reports overall results across all retrievers.

\begin{table}[h]
\centering
\caption{Retrieval benchmark results on StratRAG validation set
         ($n=200$). Best results in \textbf{bold}.}
\label{tab:results}
\begin{tabular}{lccccc}
\toprule
\textbf{Retriever} &
\textbf{R@1} & \textbf{R@2} & \textbf{R@5} &
\textbf{MRR} & \textbf{NDCG@5} \\
\midrule
Random         & 0.0525 & 0.1425 & 0.3300 & 0.3190 & 0.2336 \\
BM25           & 0.3950 & 0.6000 & 0.8150 & 0.8732 & 0.7624 \\
Dense (MiniLM) & 0.4175 & 0.6500 & 0.8600 & 0.9035 & 0.8087 \\
Hybrid         & \textbf{0.4400} & \textbf{0.6975} &
                 \textbf{0.9050} & \textbf{0.9310} &
                 \textbf{0.8543} \\
\bottomrule
\end{tabular}
\end{table}

% ── 5. Analysis & Findings ────────────────────────────────────
\section{Analysis and Findings}

\subsection{Hybrid Retrieval Consistently Outperforms Single Methods}
Hybrid fusion outperforms both BM25 and dense retrieval across all
five metrics.
The gain over dense alone is most pronounced at Recall@1
(+0.022) and Recall@5 (+0.045), suggesting that BM25's lexical
precision complements dense retrieval's semantic coverage at both
extremes of the ranking.

\subsection{Dense Retrieval Outperforms BM25 on All Metrics}
Dense retrieval (MiniLM) surpasses BM25 across all metrics, with
the largest gap at NDCG@5 (+0.046).
This indicates that semantic matching is more important than keyword
overlap for multi-hop questions, where query-document lexical
similarity is often low.

\subsection{Bridge Questions Are the Primary Open Challenge}
Table~\ref{tab:bytype} breaks down Hybrid retriever performance by
question type.

\begin{table}[h]
\centering
\caption{Hybrid retriever performance by question type
         on the StratRAG validation set.}
\label{tab:bytype}
\begin{tabular}{lrrrr}
\toprule
\textbf{Type} & \textbf{n} &
\textbf{R@2} & \textbf{MRR} & \textbf{NDCG@5} \\
\midrule
Bridge     & 171 & 0.6696 & 0.9281 & 0.8418 \\
Yes-no     &  12 & 0.8333 & 0.9167 & 0.9007 \\
Comparison &  17 & 0.8824 & 0.9706 & 0.9473 \\
\bottomrule
\end{tabular}
\end{table}

Bridge questions exhibit the lowest Recall@2 (0.67) and NDCG@5
(0.84) among all question types.
This is consistent with the nature of bridge reasoning: the
connection between the two gold documents is implicit, mediated by
an entity that appears in both but may not be present in the query
itself.
Neither BM25's lexical signals nor MiniLM's general-purpose
embeddings are well-suited to capturing this latent cross-document relationship.

Comparison questions are the easiest to retrieve (Recall@2 = 0.88),
likely because comparative queries contain strong keyword signals
(e.g., ``older'', ``larger'') that are also present in the gold
document titles.

\subsection{Recall@5 as an Upper Bound for Generation}
Hybrid Recall@5 = 0.905 indicates that, on average, 90.5\% of gold documents are retrieved within the top 5 positions. This provides a practical upper bound: a generator supplied with the top 5 retrieved documents would have access to the full supporting evidence for roughly 90\% of questions.

% ── 6. Future Work ────────────────────────────────────────────
\section{Future Work: RL-Based Retrieval for Multi-Hop Reasoning}

The benchmark results establish a clear performance ceiling for
static retrieval methods on bridge questions.
I hypothesize that this ceiling can be raised by training a
\textit{retrieval policy} using reinforcement learning, where the
reward signal is derived from downstream answer correctness rather
than query-document similarity.

Formally, let $\pi_\theta(d \mid q, \mathcal{D})$ be a retrieval
policy that selects a document $d$ from pool $\mathcal{D}$ given
query $q$.
The policy is trained to maximize expected reward:
\[
  J(\theta) = \mathbb{E}_{d \sim \pi_{\theta}(\cdot \mid q, \mathcal{D})} \left[ R(a_{\text{pred}}, a_{\text{gold}}) \right]
\]
where $R$ is a binary or soft reward based on answer correctness.

StratRAG provides the ideal evaluation foundation for this work:
fixed pool sizes, verified gold indices, and a question-type
breakdown that enables targeted analysis of where RL-based retrieval
closes the gap over hybrid fusion.

I plan to:
\begin{enumerate}
  \item Train a lightweight REINFORCE-based retrieval policy on
        the StratRAG training split
  \item Evaluate Recall@2 and MRR on the validation split,
        stratified by question type
  \item Analyze whether the policy learns to reason across
        documents for bridge questions specifically
\end{enumerate}

% ── 7. Conclusion ─────────────────────────────────────────────
\section{Conclusion}

I present StratRAG, a structured multi-hop retrieval evaluation
dataset derived from HotpotQA.
StratRAG fills a structural gap in existing benchmarks by providing
fixed-size document pools with verified gold-document indices,
enabling direct and reproducible retrieval evaluation independent of
any downstream generator.

Baseline experiments show that hybrid retrieval achieves strong
overall performance (Recall@2\,=\,0.70, MRR\,=\,0.93), but bridge
questions remain a significant open challenge (Recall@2\,=\,0.67).
This gap motivates future work on RL-based retrieval policies that learn to reason across documents using answer correctness as a reward
signal.

StratRAG is fully open-source and available at\\
\url{https://huggingface.co/datasets/Aryanp088/StratRAG}.

% ── Acknowledgements ──────────────────────────────────────────
\section*{Acknowledgements}

StratRAG is derived from the HotpotQA dataset
\citep{yang2018hotpotqa}, which was created by Zhilin Yang, Peng Qi,
Saizheng Zhang, Yoshua Bengio, William W. Cohen, Ruslan Salakhutdinov,
and Christopher D. Manning, and released under the CC BY-SA 4.0
license.
I am grateful to the HotpotQA authors for making their data openly
available to the research community.

% ── References ────────────────────────────────────────────────
\bibliographystyle{plainnat}

\end{document}